\documentclass[11pt,letterpaper]{article}

\usepackage{geometry}
\usepackage{bbm}
\usepackage[toc,page]{appendix}
\usepackage{amsmath, amsthm, amsfonts, amssymb}
\usepackage{mathrsfs} 
\usepackage{hyperref}
\usepackage{cleveref}
\usepackage{slashed}
\usepackage{dsfont}
\usepackage{wrapfig}
\usepackage{minibox}
\usepackage{graphicx}
\usepackage{graphics}
\usepackage{epsfig}
\usepackage{xcolor}
\usepackage{array}
\usepackage{tikz}
\usepackage{tikz-cd}
\usetikzlibrary{arrows,calc}
\usepackage{relsize}
\usetikzlibrary{decorations.pathmorphing,calc}
\usepackage{authblk}
\usepackage[labelfont=bf]{caption}
\usepackage{caption}
\usepackage{subcaption}
\usepackage{placeins}

\def\D{\Delta}

\def\m{\mu}

\def\l{\lambda}

\def\r{\rho}
\def\s{\sigma}
\def\thintablerule{\hrule height0.4pt}

\newcommand{\be}{\begin{equation}}
\newcommand{\ee}{\end{equation}}
\newcommand{\bea}{\begin{eqnarray}}
\newcommand{\eea}{\end{eqnarray}}
\newcommand{\bal}{\begin{aligned}}
\newcommand{\eal}{\end{aligned}}
\newcommand{\eq}[1]{Eq.~(\ref{#1})}

\numberwithin{equation}{section}

\allowdisplaybreaks

\begin{document}

\tikzset{
    photon/.style={decorate, decoration={snake}, draw=black},
    electron/.style={draw=black, postaction={decorate},
        decoration={markings,mark=at position .55 with {\arrow[draw=black]{>}}}},
    gluon/.style={decorate, draw=black,
        decoration={coil,amplitude=4pt, segment length=5pt}} 
}

\centerline{\LARGE The conformal $N$-point scalar correlator in coordinate space}
\vskip .5cm

\vskip 2 cm
\centerline{\large Nikos Irges$^1$, Fotis Koutroulis$^1$ and Dimosthenis Theofilopoulos$^2$}
\vskip 1cm
\centerline{\it 1. Department of Physics}
\centerline{\it National Technical University of Athens}
\centerline{\it Zografou Campus, GR-15780 Athens, Greece}
\centerline{2. \it Dipartimento di Matematica e Fisica}
\centerline{\it Universita del Salento and INFN Sezione di Lecce }
\centerline{\it Via Arnesano 73100, Lecce Italy}
\centerline{\it e-mail: irges@mail.ntua.gr, fkoutroulis@central.ntua.gr, dimosthenis.theofilopoulos@le.infn.it}

\vskip 2.2 true cm
\thintablerule
\vskip 2.0ex

\centerline{\bf Abstract}
We present a systematic derivation of the form of correlators of $N$ operators in a Conformal Field Theory in $d>2$ dimensions
and the exchange-symmetry constraints that the functions of the dimensionless cross-ratios obey for $N>3$.

\vskip 1.0ex\noindent
\vskip 2.0ex
\thintablerule



\pagebreak



\section{Introduction}\label{Intro}

Quantum Field Theories may run into a fixed point (or a fixed line etc.) in their phase diagram.
Then the space-time symmetry of the system may be enhanced from Poincare to Conformal symmetry. 
The most familiar case is that of a Gaussian fixed point where the theory is free in which case, if massless, it may be described by a 
free Conformal Field Theory (CFT) \cite{Slava, Slava2}. All correlators in such theories are constrained by the requirement of their invariance
under the action of the conformal group. The 2 and 3-point functions are completely fixed up to normalization, while the 4-point function is only partially constrained,
with a 2-parameter freedom remaining after all conformal Ward identities have been imposed. 

In this letter we present the computation of the correlator of $N$ scalar operators in CFT in coordinate space\footnote{An earlier attempt is \cite{Stotkov}.
There is also recent interesting actvity to express such correlators in momentum space \cite{Skenderis, Claudio, Dimos, Oh}. } and give two explicit examples.
First we rederive the 4-point function and then we give the example of the 6-point function which has not appeared before.

\section{The scalar correlator in $x$-space}

The form of the correlator of four scalar operators ${\cal O}(x_i)$ of the same scaling dimension $\D$ in $d\, (>2)$ dimensions, located at space-time points $x_i$
in a CFT is constrained by the conformal symmetry $SO(2,d)$\footnote{We will switching back and forth from Minkowski to Euclidean signature-in which case 
the space-time symmetry is $SO(1,d+1)$-depending on the situation.} to
\be\label{4cor}
\langle {\cal O}(x_1){\cal O}(x_2){\cal O}(x_3){\cal O}(x_4) \rangle = R_4\, g(u,v)\, ,\hskip 1cm R_4=\frac{1}{x_{12}^{2\D}x_{34}^{2\D}}\, ,
\ee
where $x_{ij}=|x_i-x_j|$.\footnote{We will abuse this notation and sometimes use the same notation for the vector $x_i^\m-x_j^\m$ itself.
Which is the correct reading, should be clear form the context. When $x_{ij}$ is raised to an even power the two are equivalent.}
The conformally invariant cross-ratios $u$ and $v$ are defined as
\be
u= \frac{x_{12}x_{34}}{x_{13}x_{24}}, \hskip 1cm v= \frac{x_{23}x_{14}}{x_{13}x_{24}}\, .
\ee
The function $g(u,v)$ remains unconstrained by the
conformal symmetry itself, but satisfies additional relations, obtained by the requirement that the correlator, in Euclidean space, is symmetric
under the interchange $x_i \leftrightarrow x_j$, symbolized by the notation $(i\, j)$. The action $(i\, j)$ on a function of coordinates 
induces an action denoted as $g_{ij}$. Invariance of the correlator under all possible such exchanges imposes the two exchange-symmetry constraints
\be\label{4cross}
g(u,v) = g\left(\frac{u}{v}, \frac{1}{v}\right), \hskip 1cm g(u,v) = \left(\frac{u}{v}\right)^{2\D} g(v,u)\, .
\ee
In the absence of additional input, like the Operator Product Expansion, these are (the only) independent constraints on $g$.
Recall finally that $R_4$ and $g(u,v)$ are both and separately conformally invariant, so that the correlator in \eq{4cor} 
can be seen as being factorized in coordinate space, in the product of at least two invariant substructures.

The simplest way to derive the conditions in \eq{4cross}$-$consider for simplicity a CFT in four dimensions but the generalization to arbitrary dimensions is 
straightforward$-$is to embed the system in a flat space-time of two dimensions higher \cite{Costa},
with metric of signature $(-,+,+,+,+,-)$, parametrized by the coordinates $y^A$, $A=\m,5,6$ and 
project back to the original 4d space by the null-cone condition $y^A y_A=0$ and the identification for the $4d$ coordinates
\be
x^\m = \frac{y^\m}{y^+}\, , \hskip 1cm y^+ = y_5^+ + y^+_6.
\ee
An advantage of this procedure is that the conformal transformations are just rotations and/or boosts in the 6-dimensional space and
the only non-zero invariants constructed from the coordinates are the inner products 
\be
y_i\cdot y_j = -\frac{1}{2} (y_i^+ y_j^+) (x_i-x_j)^2\, .
\ee
It can be shown that the fields ${\cal O}_q(x) = (y^+)^{\D_q} \Phi (x,y^+)$ depend only on $x$, have scaling dimension $\D_q$ and for a
conformally invariant correlator the $\Phi$'s must contribute to it terms proportional to the product of all possible inner products
$y_i\cdot y_j$. The 4-point function for example must be of the form
\be
\langle {\cal O}_1(x_1){\cal O}_2(x_2){\cal O}_3(x_3){\cal O}_4(x_4) \rangle = \frac{\prod_{a=1,\cdots,4}\left(y^+_a\right)^{\D_a}}{\prod_{i,j} \left(y_i\cdot y_j\right)^{e_{ij}}}\, ,
\hskip .75cm 1,2,3 =i < j=2,\cdots,4
\ee
Imposing the self consistency condition that the right hand side is $y_a^+$-independent and restricting to identical scalar operators,
we arrive at \eq{4cor}. By acting on the result with $g_{12}$ and $g_{13}$ and requiring invariance of the correlator, we obtain \eq{4cross}.

This methodology can be straightforwardly generalized. For the correlator of $N$ scalar operators
\be\label{Ncor}
\langle {\cal O}_1(x_1)\cdots {\cal O}_N(x_N) \rangle \sim \frac{\prod_{a=1,\cdots,N}\left(y^+_a\right)^{\D_a}}{\prod_{i,j} \left(y_i\cdot y_j\right)^{e_{ij}}}\, ,
\hskip .75cm 1,\cdots, N-1 =i < j
\ee
 the conditions that restrict its form stem from the requirement of its independence from the $y^+$'s, as before.
Clearly we have more unknowns than equations so we must decide which $e_{ij}$ to solve for.
Since we have $N$ equations, we have to pick $N$ exponents.
Any loss of generality involved in this choice will be lifted by the exchange-symmetry constraints.
A convenient choice is to solve for $e_{1i}, \, i=2,\cdots,N$ and $e_{23}$.
Defining the vectors $E=(e_{12},\cdots, e_{1N},e_{23})$, $D=(\D_1,\D_i - (\s^N_i + \r_2^i))$ the equation
to be solved for $E$ is ($T$ stands for transpose)
\be\label{system}
M E^T = D^T\, .
\ee
In the above we have defined the partial sums
\bea
\s_i^N &=& e_{i,i+1} + e_{i,i+2}+\cdots + e_{i,N} \nonumber\\
\r_2^i &=& e_{2,i}+e_{3,i}+\cdots + e_{i-1,i}
\eea
 where $i,j=2,\cdots,N$ and $\s_i^j$ and $\r_i^j $ are non-zero only when $i<j$. The matrix $M$ is 
 \be
 M = \begin{bmatrix}
1 & 1 & 1 & \cdots & 1 & 1 & 0  \\
1 & 0 & 0 & \cdots & 0 & 0 & 1  \\
0 & 1 & 0 & \cdots & 0 & 0 & 1  \\
0 & 0 & 1 & \cdots & 0 & 0 & 0  \\
. & . & . & \cdots & . & . & .  \\
. & . & . & \cdots & . &. & .  \\
. & . & . & \cdots & . & . & .  \\
0 & 0 & 0 & \cdots & 1 & 0 & 0  \\
0 & 0 & 0 & \cdots & 0 & 1 & 0  \\
\end{bmatrix}\, .
 \ee
For $N>d+2$ correlators degeneracies originating from the necessary linear dependence of some of the $y_i$
may start to arise that must be dealt with \cite{Costa}. They appear by making the above matrix have some linearly dependent rows and columns.
In this case these rows/columns must be moved to the right hand side of \eq{system}, into ${D}$. As a result,
some of the exponents in $E$ will not be independent. We will not complicate our analysis any further by such a possibility since apart from this technicality the
logic is the same as for the non-degenerate case $N\le d+2$.
The easiest way to solve this system of equations is to discard the first row and last column, which leaves an
$N-1$ dimensional unit submatrix in $M$, trivially invertible. The solution is given though in terms of $e_{23}$ due to the
missing row and column. Fortunately we can solve for $e_{23}$ separately, by combining for example the sum of all $N-1$
equations with the constraint that comes from the observation that \eq{Ncor} must be invariant under the trivial rescaling $y_i\rightarrow \l y_i$.
The result is
\be\label{e23}
2e_{23} = -\D_1 + \D_2 + \cdots + \D_N - 2(\r_2^4 + \cdots + \r_2^N)
\ee
and then the $N-1$ dimensional system of equations collapses to 
\be
2e_{1i} = 2\D_i -2 (\s^N_i + \r_2^i)\, ,
\ee
where the only thing to remember is to substitute for $e_{23}$ from \eq{e23} when it appears in either $\s^N_i $ or $\r_2^i$
which occurs twice, once in $e_{12}$ and once in $e_{13}$.
It is illuminating to show the explicit form of the final solution:
\bea
2e_{23}&=& (-\D_1+\D_2+\cdots + \D_N) -2 (\r_2^4+\cdots+\r_2^N)\nonumber\\
2e_{12} &=& 2\D_2 - (-\D_1+\D_2+\cdots + \D_N) - 2 {\hat \s}_2^N + 2(\r_2^4+\cdots+\r_2^N)\nonumber\\
2e_{13} &=& 2\D_3 - (-\D_1+\D_2+\cdots + \D_N) - 2 {\s}_3^N + 2(\r_2^4+\cdots+\r_2^N)\nonumber\\
2e_{1i} &=& 2\D_i - 2(\s_i^N + \r_2^i)\, ,\hskip .75cm i=4,\cdots, N \nonumber\\
\eea
where ${\hat \s}_2^N = \s_2^N-e_{23}$. Then 
\bea
\langle {\cal O}_1(x_1)\cdots {\cal O}_N(x_N) \rangle = \frac{1}{x_{12}^{2e_{12}}\cdots x_{1N}^{2e_{1N}} x_{23}^{2e_{23}}} \frac{1}{\prod_{ij}x_{ij}^{2e_{ij}}}
\eea
 where in the product: $i=2,\cdots, N-1$ and $j=i+1,\cdots N$ and $ij\ne 23$. 
To prepare this expression for an exchange-symmetry analysis we first define
\bea
\D_{23} &=& -\D_1+\D_2+\cdots + \D_N\nonumber\\
\D_{12} &=& 2\D_2-(-\D_1+\D_2+\cdots + \D_N)\nonumber\\
\D_{13} &=& 2\D_3-(-\D_1+\D_2+\cdots + \D_N)\nonumber\\
\D_{1i} &=& 2\D_i\, , \hskip .75cm i=4,\cdots,N
\eea
and write the correlator as
\bea\label{Ncor2}
\langle {\cal O}_1(x_1)\cdots {\cal O}_N(x_N) \rangle = R_N\, 
\frac{x_{23}^{2 (\r_2^4+\cdots+\r_2^N)} \prod_{i=4}^N x_{1i}^{2(\s_i^N + \r_2^i)} }{x_{12}^{- 2 {\hat \s}_2^N + 2(\r_2^4+\cdots+\r_2^N)}x_{13}^{- 2 {\s}_3^N + 2(\r_2^4+\cdots+\r_2^N)}   
\prod_{ij}x_{ij}^{2e_{ij}}}
\eea
with the same restrictions on the $i,j$ indices as above and
\be
R_N = \frac{1}{x_{23}^{\D_{23}}\prod_{a=2}^{N} x_{1a}^{\D_{1a}}}\, .
\ee
The geometric interpretation says that if we think of the correlator as a sort of a representation of a discrete metric on the points $\{x_1,\cdots,x_N\}$ then 
$R_N$ is its radial part, the rest is the angular part and rotations correspond to transformations that exchange $x_i\leftrightarrow x_j$. 
When the $N$ operators are distinct, the gauging of the exchange-symmetry group does not leave the triangle defined by any three points invariant (apart from the identity action)
and the radial part $R_N$ will contain a $123$ sector, corresponding to the triangle defined by $x_1,x_2,x_3$. For $N=3$ this is a conformally invariant structure.
When the operators in the correlator are identical, it may happen that a non-trivial (not an identity) combination of the exchange-symmetry group elements
leaves the $123$ triangle invariant and then the corresponding sector has no reason to appear in $R_N$. 
Instead, all information for structures built from triangles is contained in the angular part $f$. Such is the case of the $N=4$ correlator of identical scalars.

The statement of exchange-symmetry (in Euclidean $x$-space) is that 
\be
g_{1a} \langle {\cal O}_{1}(x_1)\cdots {\cal O}_{N}(x_N) \rangle = \langle {\cal O}_{1}(x_1)\cdots {\cal O}_{N}(x_N) \rangle\, , \hskip 1cm a=2,\cdots,N
\ee
since the $(1\, a)$ generate the permutation group $S_N$. We also define here the important quantity
\be
J_a = R_N^{-1}\, (g_{1a} R_N)\, ,
\ee
a sort of discrete version of a Jacobian, originating from the transformation induced by the $g_{1a}$.
The only ingredient we are missing are the conformally invariant cross-ratios. These can be straightforwardly obtained from
\eq{Ncor2} by collecting all the $x_{mn}$ under a fixed power $2e_{kl}$. This is not a unique decomposition of the correlator but is easy to generalize. Then, we obtain
the $2(N-3)$ conformally invariant, order two, cross-ratios
\be\label{u2}
u_{2k} = \frac{x_{23}x_{1k}}{x_{13}x_{2k}}\, ,\hskip 1cm u_{3k} = \frac{x_{23}x_{1k}}{x_{12}x_{3k}}\, , \hskip .5 cm k=4,\cdots,N
\ee
and the $\frac{1}{2}(N-3)(N-4)$, order three, cross-ratios\footnote{The existence of these has been noticed in \cite{Rosenhaus} for $N=5$.}
\be\label{u3}
u_{ji} = \frac{x_{23} x_{1i}x_{1j}}{x_{12}x_{13}x_{ji}}\, , \hskip 1cm 4\le j < i=5,\cdots,N
\ee
These are ratios of 3-point functions but being dimensionless moduli, appear in the angular part of the correlator.
The counting is right, since $1+2+\cdots + (N-4) + 2(N-3) = \frac{1}{2}N(N-3)$.
It seems that cross-ratios of higher order do not form and any higher order cross-ratio can be expressed in terms of the order two
and the order three ratios, may it be of even or odd order. An interesting fact is that while for $N=4$ we see only order two cross-ratios
and for $N=6$ the order two are twice as many as the order three ones, for large $N$ the order three cross ratios start to dominate.
Note also the useful identities
\be\label{permgroupu}
g_{23} u_{2k} = u_{3k} \, ,\hskip .5cm g_{jk} u_{ji} = u_{ki}\,\,\, (k\ne i), \hskip .5 cm g_{ik} u_{ji} = u_{jk}\,\,\, (k\ne j)
\ee
which tell us that we can start from $u_{24}$ and $u_{45}$ and generate all other cross-ratios by acting on them with the elements of $S_N$.
The last step is to generalize in the expression for the correlator the part that depends on the unfixed exponents to a general
function of its conformally invariant cross-ratios, which we will refer to also as the conformal coordinates: 
\be
\langle {\cal O}_{1}(x_1)\cdots {\cal O}_{N}(x_N) \rangle = R_N f_{1\cdots N}(u_{24},\cdots,u_{2N},u_{34},\cdots,u_{4N},\cdots,u_{N-1,N})\, .
\ee
Now we are done, since one can use these expressions and obtain explicitly all $N$-correlators of scalar operators of scaling dimension $\D_i$ from \eq{Ncor2} and their 
$N-1$ cross symmetry constraints ($a=2,\cdots,N$):
\bea\label{crossgen}
&& f_{q_1\cdots q_N}(u_{24},\cdots,u_{2N},u_{34},\cdots,u_{3N},\cdots,u_{N-1,N}) = \nonumber\\
&& J_a f_{g_{1a}[q_1\cdots q_N]}(g_{1a}u_{24},\cdots,g_{1a}u_{2N},g_{1a}u_{34},\cdots,g_{1a}u_{3N},\cdots,g_{1a}u_{N-1,N})\, .
\eea
To illustrate the general process we give two examples. We first rederive the $N=4$ correlator and then present the $N=6$ correlator for the simple case 
of identical operators, in which case $\D_i=\D$ and $f_{q_1\cdots q_N}=f$.

\subsection{The $N=4$ correlator}

For $N=4$ there are two coordinates of the type \eq{u2}:
\be
u_{24} = \frac{x_{23}x_{14}}{x_{13}x_{24}}\, , \hskip 1cm u_{34} = \frac{x_{23}x_{14}}{x_{12}x_{34}}
\ee
and no coordinates of the type \eq{u3}. Also, for identical operators $\D_{12}=\D_{13}=0$ and $\D_{14}=\D_{23}=2\D$.
The correlator in this case is
\be
\langle {\cal O}(x_1){\cal O}(x_2){\cal O}(x_3){\cal O}(x_4) \rangle = R_4 f(u_{24},u_{34})\, , \hskip .5cm R_4=\frac{1}{x_{14}^{2\D}x_{23}^{2\D}}
\ee
The action of the generators of $S_4$ on the conformal coordinates are
\bea
&& g_{12} u_{24} = \frac{1}{u_{24}}\, , \hskip .75cm g_{12} u_{34} = \frac{u_{34}}{u_{{24}}}\nonumber\\
&& g_{13} u_{24} = \frac{u_{24}}{u_{34}}\, , \hskip .75cm g_{13} u_{34} = \frac{1}{u_{{34}}}\nonumber\\
&& g_{14} u_{24} = u_{34}\, , \hskip .85cm g_{14} u_{34} = u_{24}
\eea
and the three Jacobians are
\be
J_2 = u_{24}^{2\D}\, , \hskip .75cm J_3= u_{34}^{2\D}\, , \hskip .75cm  J_4=1\, .
\ee
These imply the three exchange-symmetry constraints
\bea\label{4crosscov}
g_{12}:\hskip .25cm f(u_{24},u_{34}) &=& u_{24}^{2\D} f\left(\frac{1}{u_{24}},\frac{u_{34}}{u_{24}}\right)\nonumber\\
g_{13}:\hskip .25cm f(u_{24},u_{34}) &=& u_{34}^{2\D} f\left(\frac{u_{24}}{u_{34}},\frac{1}{u_{34}}\right)\nonumber\\
g_{14}:\hskip .25cm f(u_{24},u_{34}) &=& f(u_{34},u_{24})
\eea
We should make three comments here.
One is related to the observation that in \eq{4cross} we presented only two exchange-symmetry constraints and here we just found three.
What happens is that out of the three covariant constraints in \eq{4crosscov} only two are independent,
as it is easy to check that $g_{12}g_{13}g_{12}\sim g_{14}$, where the $\sim$ sign indicates not a group theory relation between $S_N$ elements but an equivalence of their action on 
the correlator and the $u_{24}, u_{34}$. In other words, the transformation with the unit Jacobian in \eq{4crosscov} for example is not independent.
This seems to be a reflection of the fact that one can bring the four points $x_1,x_2,x_3,x_4$ on a plane by conformal transformations, thus the trivial Jacobian.
Furthermore, one can place these points on the corners of a tilted rectangle that the gauging of the exchange-symmetry group turns into a square.
As a result, the exchange symmetry effectively reduces to the dihedral group $D_4$
and if the freedom to choose which three points define the plane on which the fourth point is projected on is taken into account,
the symmetry reduces further to $D_3$, which is isomorphic to $S_3$. The latter is generated by $g_{12}$ and $g_{13}$ indeed.
Thus, the $g_{14}$ operation can not be independent.
The second comment is that according to our previous geometric arguments we expect to see no 3-point subsector in $R_4$ as
the information about the invariance of the triangles inside the parallelogram under rotations about its two diagonals, are contained in the action of the $g_{1a}$.
Indeed, we saw that $\D_{12}=\D_{13}=0$ and the radial part of the correlator $R_4$ contains only two disconnected $x_{ij}$'s
($x_{14}x_{23}$ in the $u_{24},u_{34}$ angular coordinates and $x_{12}x_{34}$ in the $u,v$ coordinates).
The third comment is that the two independent constraints in \eq{4crosscov} are equivalent to the ones in \eq{4cross} by a coordinate change,
even though the trivial Jacobian transformation in the $(u_{24},u_{34})$ coordinates maps to a non-trivial one in the $(u,v)$ coordinates and vice versa.

\subsection{The $N=6$ correlator}

Let us look at a slightly more complicated example, the one of the conformal correlator of six scalar operators. 
The algorithm we described then yields via \eq{u2} and \eq{u3} the nine ($=3+3+2+1$) conformal coordinates
\bea
&& u_{24} = \frac{x_{14}x_{23}}{x_{13}x_{24}},\hskip .75cm u_{25} = \frac{x_{15}x_{23}}{x_{13}x_{25}},\hskip .75cm u_{26} = \frac{x_{16}x_{23}}{x_{13}x_{26}}\nonumber\\
&& u_{34} = \frac{x_{14}x_{23}}{x_{12}x_{34}},\hskip .75cm u_{35} = \frac{x_{15}x_{23}}{x_{12}x_{35}},\hskip .75cm u_{36} = \frac{x_{16}x_{23}}{x_{12}x_{36}}\nonumber\\
&& \hskip 3.15cm u_{45} = \frac{x_{14}x_{15}x_{23}}{x_{12}x_{13}x_{45}},\hskip .25cm u_{46} = \frac{x_{14}x_{16}x_{23}}{x_{12}x_{13}x_{46}}\nonumber\\
&& \hskip 6.35cm u_{56} = \frac{x_{15}x_{16}x_{23}}{x_{12}x_{13}x_{56}}
\eea
in terms of which the 6-point correlator is
\be
\langle {\cal O}_1(x_1)\cdots {\cal O}_6(x_6)\rangle = R_6 f_{q_1\cdots q_6}\left(u_{24},u_{25},u_{26},u_{34},u_{35},u_{36},u_{45},u_{46},u_{56}\right)
 \ee
 with the radial prefactor
 \be
  R_6 = \frac{1}{x_{12}^{\D_{12}}x_{13}^{\D_{13}}x_{14}^{\D_{14}}x_{15}^{\D_{15}}x_{16}^{\D_{16}}x_{23}^{\D_{23}}}\, .
 \ee
 In the case of scalar operators of the same scaling dimensions $\D$ this reduces to
 \be
 \langle {\cal O}(x_1)\cdots {\cal O}(x_6)\rangle = \left(\frac{x_{12}x_{13}}{x_{14}x_{15}x_{16}x_{23}^2}\right)^{2\D}\, f\left(u_{24},u_{25},u_{26},u_{34},u_{35},u_{36},u_{45},u_{46},u_{56}\right)\, .
\ee
The five corresponding exchange-symmetry constraints are
\bea
g_{12}: \,\, && f\left(u_{24},u_{25},u_{26},u_{34},u_{35},u_{36},u_{45},u_{46},u_{56}\right)=\nonumber\\
&& (u_{24}u_{25}u_{26})^{2\D}
f\left(\frac{1}{u_{24}},\frac{1}{u_{25}},\frac{1}{u_{26}},\frac{u_{34}}{u_{24}},\frac{u_{35}}{u_{25}},\frac{u_{36}}{u_{26}},\frac{u_{45}}{u_{24}u_{25}},\frac{u_{46}}{u_{24}u_{26}},\frac{u_{56}}{u_{25}u_{26}}\right)
\eea
\bea
g_{13}: \,\, && f\left(u_{24},u_{25},u_{26},u_{34},u_{35},u_{36},u_{45},u_{46},u_{56}\right)=\nonumber\\
&& (u_{34}u_{35}u_{36})^{2\D}
f\left(\frac{u_{24}}{u_{34}},\frac{u_{25}}{u_{35}},\frac{u_{26}}{u_{36}},\frac{1}{u_{34}},\frac{1}{u_{35}},\frac{1}{u_{36}},\frac{u_{45}}{u_{34}u_{35}},\frac{u_{46}}{u_{34}u_{36}},\frac{u_{56}}{u_{35}u_{36}}\right)
\eea
\bea
g_{14}: \,\, && f\left(u_{24},u_{25},u_{26},u_{34},u_{35},u_{36},u_{45},u_{46},u_{56}\right)=\nonumber\\
&& \left(\frac{u_{45}u_{46}}{u_{24}u_{34}}\right)^{2\D}
f\left(u_{34},\frac{u_{25}u_{34}}{u_{45}},\frac{u_{26}u_{34}}{u_{46}},u_{24},\frac{u_{24}u_{35}}{u_{45}},\frac{u_{24}u_{36}}{u_{46}},\frac{u_{24}u_{34}}{u_{45}},\frac{u_{24}u_{34}}{u_{46}},\frac{u_{24}u_{34}u_{56}}{u_{45}u_{46}}\right)\nonumber\\
\eea
\bea
g_{15}: \,\, && f\left(u_{24},u_{25},u_{26},u_{34},u_{35},u_{36},u_{45},u_{46},u_{56}\right)=\nonumber\\
&& \left(\frac{u_{45}u_{56}}{u_{25}u_{34}}\right)^{2\D}
f\left(\frac{u_{24}u_{35}}{u_{45}},u_{35},\frac{u_{26}u_{35}}{u_{56}},\frac{u_{25}u_{34}}{u_{45}},u_{25},\frac{u_{25}u_{36}}{u_{56}},\frac{u_{25}u_{35}}{u_{45}},\frac{u_{25}u_{35}u_{46}}{u_{45}u_{56}},\frac{u_{25}u_{35}}{u_{56}}\right)\nonumber\\
\eea
\bea
g_{16}: \,\, && f\left(u_{24},u_{25},u_{26},u_{34},u_{35},u_{36},u_{45},u_{46},u_{56}\right)=\nonumber\\
&& \left(\frac{u_{46}u_{56}}{u_{26}u_{36}}\right)^{2\D}
f\left(\frac{u_{24}u_{36}}{u_{46}},\frac{u_{25}u_{36}}{u_{56}},u_{36},\frac{u_{26}u_{34}}{u_{46}},\frac{u_{26}u_{35}}{u_{56}},u_{26},\frac{u_{26}u_{36}u_{45}}{u_{46}u_{56}},\frac{u_{26}u_{36}}{u_{46}},\frac{u_{26}u_{36}}{u_{56}}\right)\nonumber\\
\eea
There is no trivial Jacobian, so we expect these constraints to be independent. 

\section{Conclusion}

We computed the correlator of $N$ scalar operators in CFT, in coordinate space and gave two explicit examples,
for $N=4$ and $N=6$. We found that in the $N=6$ case there appear order three conformally invariant cross-ratios of six $x_{ij}$'s,
in addition to the well known order two cross-ratios of the $N=4$ case. We also gave the corresponding exchange-symmetry 
constraints associated with correlators of $N$ scalar operators with $N\ge 4$.






{\bf Acknowledgements}

We would like to thank A. Kalogirou and S. Kastrinakis for discussions.


\begin{thebibliography}{9}

\bibitem{Slava}
See for example the recent review,
S. Rychkov, arXiv:1601.05000 [hep-th].

\bibitem{Slava2}
D. Poland, S.Rychkov and A. Vichi, Rev. Mod. Phys. {\bf 91} (2019) 015002,  arXiv:1805.04405 [hep-th].

\bibitem{Stotkov}
G. M. Sotkov and R. P. Zaikov, Rept. Math.Phys. {\bf 19}  (1984) 335.

\bibitem{Skenderis}
A. Bzowski, P. McFadden and K. Skenderis,
arXiv:1910.10162 [hep-th].

\bibitem{Claudio}
C. Corian\'o, M. M. Maglio, JHEP {\bf 1909} (2019), 107, arXiv:1903.05047 [hep-th].

\bibitem{Dimos}
C. Corian\'o, M. M. Maglio and D. Theofilopoulos, arXiv:1912.01907 [hep-th].
\bibitem{Oh}
J.-H. Oh,
arXiv:2001.05379 [hep-th].

\bibitem{Costa}
M. S. Costa, J. Penedones, D. Poland and S. Rychkov,
JHEP {\bf 1111} (2011) 071,  arXiv:1107.3554 [hep-th].

\bibitem{Rosenhaus}
V. Rosenhaus, JHEP {\bf 1902} (2019) 142, arXiv:1810.03244 [hep-th].
J-F. Fortin and W. Skiba, arXiv:1905.00036 [hep-th].

\end{thebibliography}
\end{document}